\documentclass[graybox]{svmult}

\usepackage{type1cm}        
\usepackage{makeidx}         
\usepackage{graphicx}        
\usepackage{multicol}        
\usepackage[bottom]{footmisc}

\usepackage{amsfonts}
\usepackage{times}
\usepackage{amsmath}
\usepackage{graphicx}
\usepackage{amssymb}
\usepackage{xcolor,comment}
\usepackage{url}
\begin{document}

\title*{Breathers in the fractional Frenkel-Kontorova model}
\author{J. Catarecha, J. Cuevas-Maraver and P.G. Kevrekidis}

\institute{J. Catarecha
\at
Grupo de Fisica No Lineal,\\
Departamento de Fisica Aplicada I, Universidad de Sevilla,\\
Escuela Polit\'ecnica Superior, C/ Virgen de \'Africa, 7\\
41011 Sevilla, Spain\\
\email{catarechajorge@gmail.com}
\and
Jes\'us Cuevas-Maraver
\at
Grupo de Fisica No Lineal,\\
Departamento de Fisica Aplicada I, Universidad de Sevilla,\\
Escuela Polit\'ecnica Superior, C/ Virgen de \'Africa, 7\\
41011 Sevilla, Spain
\and
Jes\'us Cuevas-Maraver
\at
Instituto de Matem\'aticas de la Universidad de Sevilla (IMUS)\\
Edificio Celestino Mutis,\\
Avda. Reina Mercedes s/n\\
41012 Sevilla, Spain\\
\email{jcuevas@us.es}
\and
Panayotis G. Kevrekidis
\at
Department of Mathematics and Statistics,\\
University of Massachusetts,\\
Amherst, MA 01003-4515, USA\\
\email{kevrekid@umass.edu}
}

\maketitle

\abstract{In the present chapter, we explore the possibility of
a Frenkel-Kontorova (discrete sine-Gordon) model to
bear interactions that decay algebraically with space,
inspired by the continuum limit of the corresponding
fractional derivative. In such a setting, we revisit
the realm of discrete breathers including onsite, inter-site
and out-of-phase ones and identify their power-law spatial
decay, as well as explore their corresponding stability analysis,
by means of Floquet multipliers. The relevant stability is
also explored parametrically as a function of the frequency
and connected to stability criteria for breather dependence
of energy vs. frequency. Finally, by suitably perturbing
the breathers, we also generate moving waveforms and explore
their radiation and potential robustness.}

\section{Introduction}

The study of fractional differential equations (and fractional
calculus more broadly) has been a topic that has gained
considerable traction over the past few years, both in
mathematics~\cite{m3}, as well as in physics~\cite{m1}.
Indeed, from the point of view of applications, there
have been intriguing connections proposed to areas
such as plasma physics~\cite{m4}, cardiac electrical
propagation~\cite{m5} and even quantum mechanics~\cite{m6,m7}.
In the same spirit, a proposal that captured a considerable
amount of attention was that of~\cite{m8}. This work leveraged the
connection of optics to the Schr{\"o}dinger equation,
to propose a realization of such a fractional model
in the description of
transverse light dynamics in aspherical optical cavities.
Indeed, recently the domain of optics has offered
an experimental realization of such a fractional model
in the context of the observation of femtosecond
laser pulses in the temporal domain~\cite{m8b}.

While the description of continuum fractional media
has been gaining considerable traction, more recently
discrete variants of fractional models have also
been proposed and studied. Prototypical among them
has been both the one-dimensional~\cite{m9} and
two-dimensional~\cite{m10} discrete nonlinear
Schr{\"o}dinger equation, following the earlier
rigorous mathematical work of~\cite{m13}, connecting
continuum and discrete such models; see also the
recent work of~\cite{aceves}. The relevant studies have
been extended to the examination of fractional
nonlinear electrical lattices~\cite{m11} and the consideration
of impurities in such fractional lattice models, see for instance~\cite{m12}.

One of the most prototypical discrete models in connection
to the study of nonlinear excitations is, arguably, the
discrete sine-Gordon equation, or as it is otherwise
known the Frenkel-Kontorova model~\cite{braun1998,braun2004}.
While it was originally proposed in the late 1930's as
a model for the propagation of a crystal lattice
in the vicinity of a dislocation core, this model has
reappeared in a variety of different physical realizations.
Arguably, the simplest (and most experimentally accessible)
realization thereof is that of a chain of coupled (via elastic springs)
torsion pendula~\cite{scott,english}. However, there are
numerous other ones including the technologically
relevant realm of superconducting Josephson junction
arrays~\cite{ustinov,floria} and the more speculative,
but interesting, in its biophysical connections,
setting of DNA dynamics; see, e.g.~\cite{yomosa,yakush}.

The aim of the present work is to combine these areas of
study. In particular, a major theme of interest within
the discrete sine-Gordon equation~\cite{floyd}
has been the analysis
of its discrete breathers, i.e., temporally periodic, spatially
localized waveforms that arise generically in nonlinear
dynamical lattices~\cite{flach,Aubry}. Our intention here
is to explore a variety of possible discrete breathers
in the context of the discrete analogue of the fractional
sine-Gordon model. We
start by identifying the relevant time-periodic states
and by visualizing their tails in order to compare/contrast
the fractional/long-range case lattice
with the standard exponential decay of regular breathers in typical
nearest-neighbor coupled lattices.
In parallel to the existence of the states, we explore
their dynamical stability by means of computing the
Floquet multipliers of the linearization around the relevant
periodic orbits. The associated instabilities are parametrically
explored as a function of the strength of the lattice
coupling $C$, as well as upon varying the frequency of the
breathers $\omega_b$. The latter is intended to explore
the connection of potential instabilities with changes
in the monotonicity of the energy vs. the frequency
of the breathers, in line with the theoretical arguments
of~\cite{ourprl}. Finally, the breathers are provided with
a ``kick'' in order to set them to motion and direct numerical
simulations serve to illustrate their potential robustness
as dynamically evolving wave states within the lattice.

Our presentation is structured as follows. In Sect. 2, we provide
the mathematical setup of the model. In Sect. 3, we
discuss our numerical findings. Finally, Sect. 4 offers
some conclusions, as well as directions for future study.

\section{Model setup}

Our starting point will be a modified Frenkel-Kontorova model
of the form:

\begin{equation}
    \label{eq:FK}
    \ddot{u}_n+\sin u_n-C\Delta_n^\alpha u_n=0,
\end{equation}
where $\Delta^\alpha$ is the discrete analogue of the fractional Laplacian given by \cite{Ciaurri}
\begin{equation}
    \label{eq:Laplacian}
    (-\Delta_n)^\alpha u_n = \sum_{m\neq n} K^\alpha(n-m)(u_n-u_m),
\end{equation}
and the kernel $K^\alpha(m)$ takes the form:
\begin{equation}
    \label{eq:kernel}
    K^\alpha(m) = \frac{4^\alpha\Gamma(\alpha+1/2)}{\sqrt{\pi}|\Gamma(-\alpha)|}\frac{\Gamma(|m|-\alpha)}{\Gamma(|m|+1+\alpha)}.
\end{equation}

Recall that in the case of $\alpha=1$,
{$K^{\alpha=1}(m)=\delta_{|m|,1}$} and
the relevant expression retrieves
the standard discrete Laplacian case which has been
previously explored at length~\cite{braun1998,braun2004,floyd}, including
as concerns how its discrete breathers approach
the corresponding continuum limit~\cite{Martina}.

With the above definition in mind, (\ref{eq:FK}) can be alternatively written as:

\begin{equation}
    \label{eq:dyn}
    \ddot{u}_n+\sin u_n+C\sum_{m>0} K^\alpha(m)(2u_n-u_{n+m}+u_{n-m})=0,
\end{equation}
which can be derived from the following Hamiltonian:

\begin{equation}
    \label{eq:ham}
    H=\sum_n E_n,\qquad E_n=\frac{\dot{u}^2_n}{2}+(1-\cos(u_n))+\frac{C}{4}\sum_{m>0}K^\alpha(m)[(u_n-u_{n+m})^2+(u_n-u_{n-m})^2]
\end{equation}

By virtue of MacKay-Aubry's theorem~\cite{MacKay}, this equation possesses discrete breather solutions, namely time-periodic, spatially localized discrete standing waves of frequency $\omega$, as long as $C\rightarrow0$, namely, at the anti-continuum limit. In order to find such solutions, one must solve the dynamical equation (\ref{eq:dyn}) at $C=0$
(i.e., for isolated oscillators) and subsequently continue
the identified oscillation to the desired coupling.
Indeed, here it is assumed that one or few oscillators
are initiated in this oscillatory state while the rest
are at rest for the $C=0$ limit.
As is well-known, (\ref{eq:dyn}) transforms into the
sine-Gordon equation in the long-wavelength
limit, i.e., when $C\rightarrow\infty$. Interestingly,
this is the only case of a Klein-Gordon type (spatially
homogeneous) partial differential equation possessing breathers at the continuum limit~\cite{birnir}.
Naturally, this is also associated with the integrability
of the continuum model which mitigates the resonances between
the breather frequency and the modes of the continuous spectrum of the problem.
On the
discrete side, equation (\ref{eq:dyn}) can be cast as a Klein-Gordon oscillator chain with long-range interaction, in which breathers have been previously studied in \cite{Fflach,JesusDNA}. More recently, such long-range
lattice problems were started to be explored in
experimental settings, as seen, e.g., in the work
of~\cite{chon}, although we are not aware at the moment of
a realization thereof, involving, e.g., the setting
of torsion pendula (or similar) of interest to the sine-Gordon context.

One of the ways for solving (\ref{eq:dyn}) is to express the (time-reversible) solution as a truncated Fourier cosine series:

\begin{equation}
    \label{eq:Fourier}
    u_n(t)=z_{0,n}+2\sum_{k=1}^{k_m}z_{k,n}\cos(k\omega t)
\end{equation}
so that the set of ODEs transforms into a set of nonlinear algebraic equations for the expansion coefficients
$z_{k,n}$ that can be solved by fixed point methods, such as
the Newton-Raphson (see e.g. \cite{Marin} for more details).

Stability properties of breather solutions are found by means of Floquet analysis; see \cite{Aubry} for a detailed
exposition {and \cite{ourphi4} for a review}. To this aim, a perturbation $\xi_n$ is added to the solution at (\ref{eq:dyn}), leading to the following equation

\begin{equation}
    \ddot{\xi}_n+\cos (u_n)\xi_n+C\sum_{m\geq1} K^\alpha(m)(2\xi_n-\xi_{n+m}-\xi_{n-m})=0
\end{equation}

In order to perform the Floquet analysis, one must then compute the spectrum of the Floquet operator, whose matrix representation is known as the monodromy matrix, and is defined from the following map:

\begin{equation}
    \Xi(T)=\mathcal{F}\Xi(0),\qquad \Xi(t)=[{\xi_n(t)},{\dot{\xi}_n(t)}]
\end{equation}

The eigenvalues of $\mathcal{F}$ are known as Floquet multipliers, and can be expressed in the form $\lambda=\exp(i\theta)$. As the Floquet operator is real and symplectic, the multipliers come in pairs $(\lambda,1/\lambda)$ if they are real, or in quadruplets $(\lambda,\lambda^*,1/\lambda,1/\lambda^*)$ if they are complex. For a periodic solution to be stable, it is needed that $|\lambda|\leq1$. As our system is Hamiltonian, for the purposes of stability, it is needed that all the eigenvalues are on the unit circle. Additionally, there is always a pair of eigenvalues at $\theta=0$, given its
Hamiltonian nature.

An interesting issue of the fractional Klein-Gordon models is the form of the linear modes (phonon) band. These modes, of the plane wave form $u_n=u_0\exp(i(qn-\Omega t))$, fulfill the dispersion relation

\begin{equation}
    \label{eq:phonons1}
    \Omega(q)^2=1+4C\sum_{m\geq1}K^\alpha(m)\sin^2\left(\frac{mq}{2}\right)
\end{equation}

{The summation can be written in a closed form by using the Gauss series (15.1.1) in \cite{NIST}}

\begin{equation}
    \label{eq:phonons}
    \Omega^2(q)=1+2C\frac{\Gamma(2\alpha)}{\Gamma(1+\alpha)\Gamma(\alpha)}
    \left[1-\left[F(1,-\alpha,1+\alpha;\mathrm{e}^{-iq})+F(1,-\alpha,1+\alpha;\mathrm{e}^{iq})\right]\right]
\end{equation}
with $F(a,b,c;z)\equiv{}_2F_1(a,b,c;z)$ being the hypergeometric function. From (\ref{eq:phonons1}), it is easy to see that the phonon band is contained between $\Omega(0)$ and $\Omega(\pi)$, i.e., the bottom (top) of the phonon band correspond to the $q=0$ ($q=\pi$) phonon. Consequently, the phonon spectrum for our computations below is
expected to lie in the range:

\begin{equation}
    \label{eq:phononband}
    \Omega^2_\mathrm{ph}\in\left[1,1+2C\frac{\Gamma(2\alpha)}{\Gamma(1+\alpha)\Gamma(\alpha)}
    \left[2F(1,-\alpha,1+\alpha;-1)-1\right]\right]
\end{equation}

\begin{figure}
\begin{center}
\includegraphics[width=.75\textwidth]{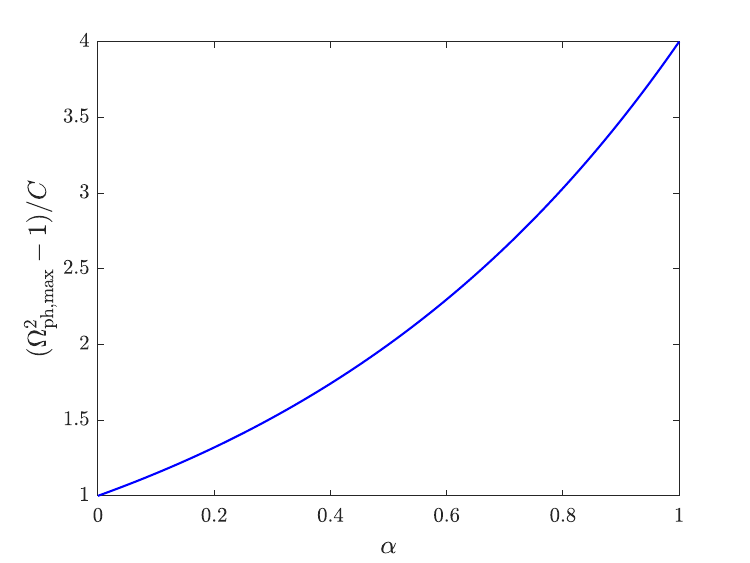}
\end{center}
\caption{Normalized distance between the lower and upper
band edges for the long-range sine-Gordon lattice, represented
as a function of the fractionality parameter $\alpha$.}
\label{fig:phonon}
\end{figure}

{Figure~\ref{fig:phonon} shows the dependence of $(\Omega^2_\mathrm{ph,max}-1)/C$, with $\Omega_\mathrm{ph,max}$ being the value of the upper edge of the phonon band, with respect to $\alpha$. This value tends to $4$ when $\alpha\rightarrow1$, as expected, whereas it tends to a flat band when $\alpha\rightarrow0$.}

\section{Numerical analysis}

In this section we will present the main findings regarding the continuation of discrete breathers from the anti-continuum to the continuum limit, in a similar fashion to what is done in \cite{Martina}; we will also explore the variations of the breather branches upon
changes in frequency. In particular, we will focus on the bifurcations emerging in this process and how their features depend on the fractionality, as parametrized by $\alpha$. {In addition, we will restrict to the $0.5\leq\alpha\leq1$ regime, as for smaller values of $\alpha$, the breather decays so slowly that a very large chain must be considered; otherwise, the value of $u_n$ at the boundaries
is  far from being well approximated as vanishing, creating artificial inhomogeneities therein.
It is relevant to point out that other Chapters
in the present volume (such as the
one on ``Numerical methods for fractional PDEs''
by C. Klein and N. Stoilov) address the
technical challenges of such low values of the
fractional exponent, and the need for
specialized multi-domain approaches therein. We
defer to these studies for considerations
on how to address the latter regime of $\alpha$'s.}

\begin{figure}
\begin{center}
\begin{tabular}{ccc}
\includegraphics[width=.33\textwidth]{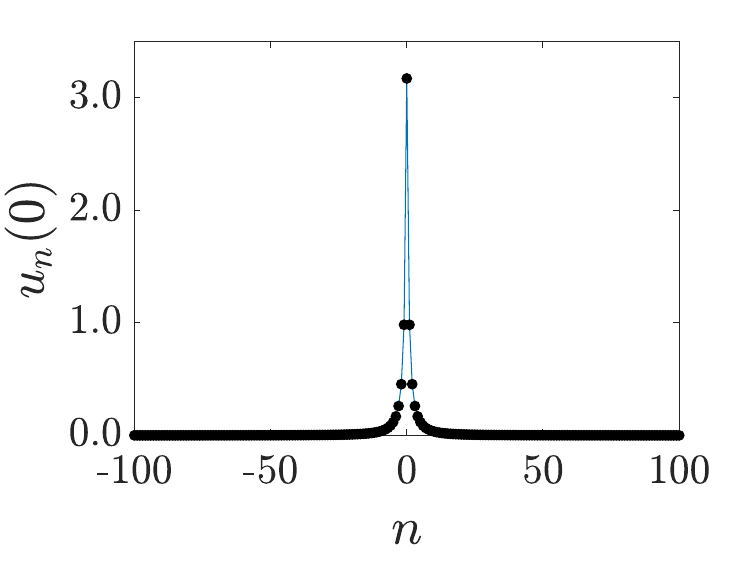} &
\includegraphics[width=.33\textwidth]{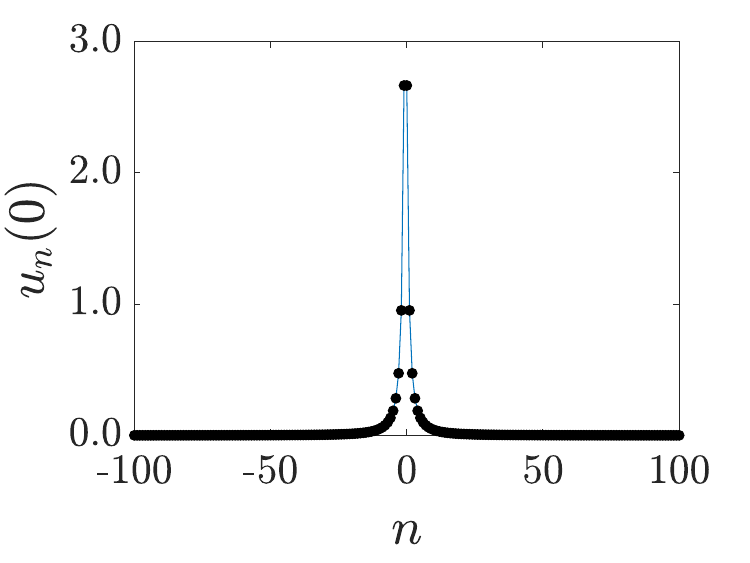} &
\includegraphics[width=.33\textwidth]{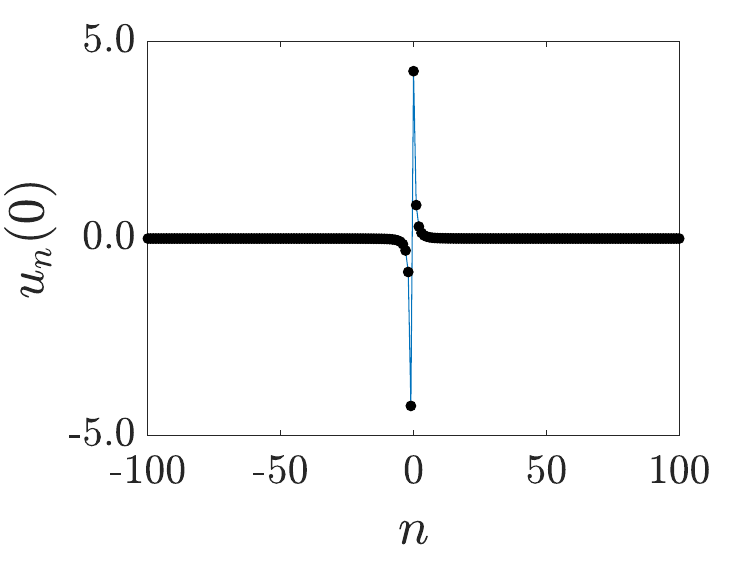} \\
\includegraphics[width=.33\textwidth]{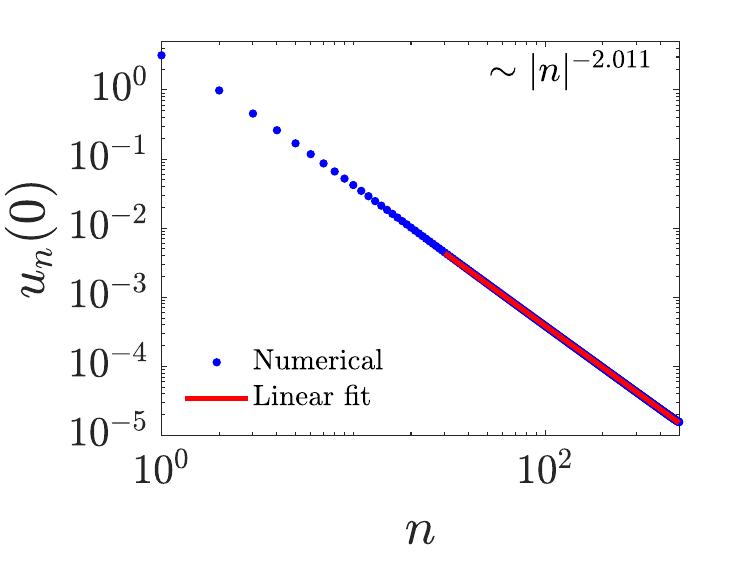} &
\includegraphics[width=.33\textwidth]{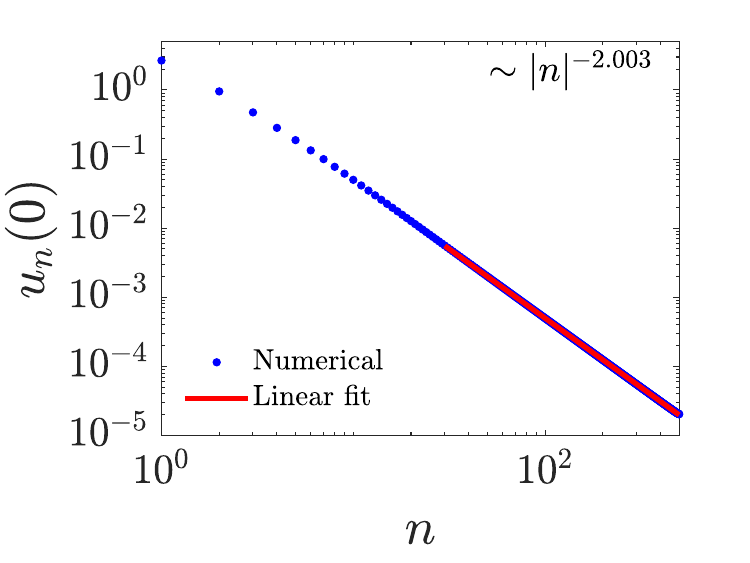} &
\includegraphics[width=.33\textwidth]{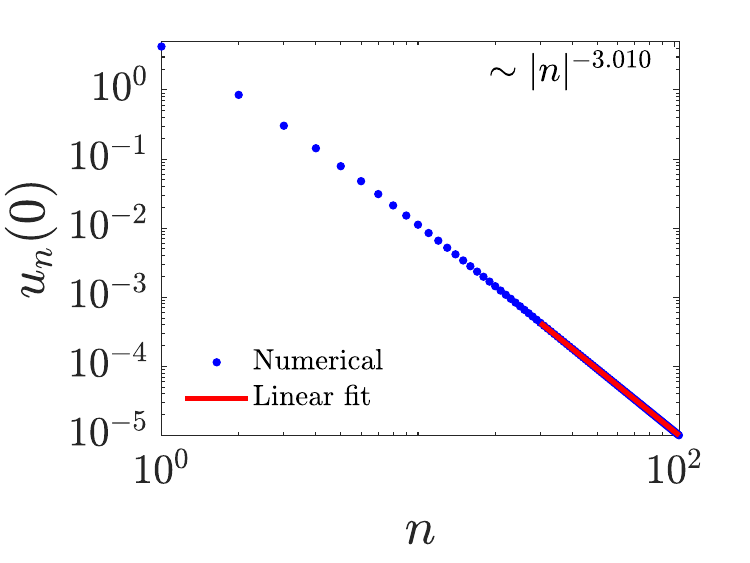} \\
\includegraphics[width=.33\textwidth]{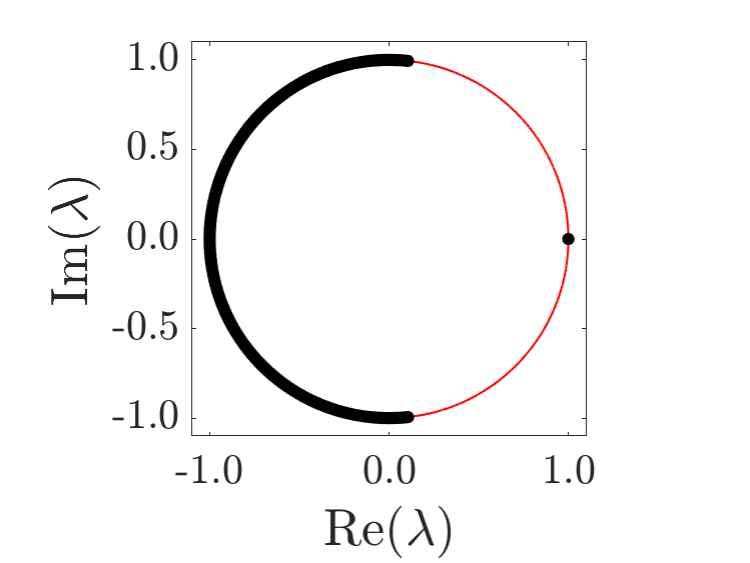} &
\includegraphics[width=.33\textwidth]{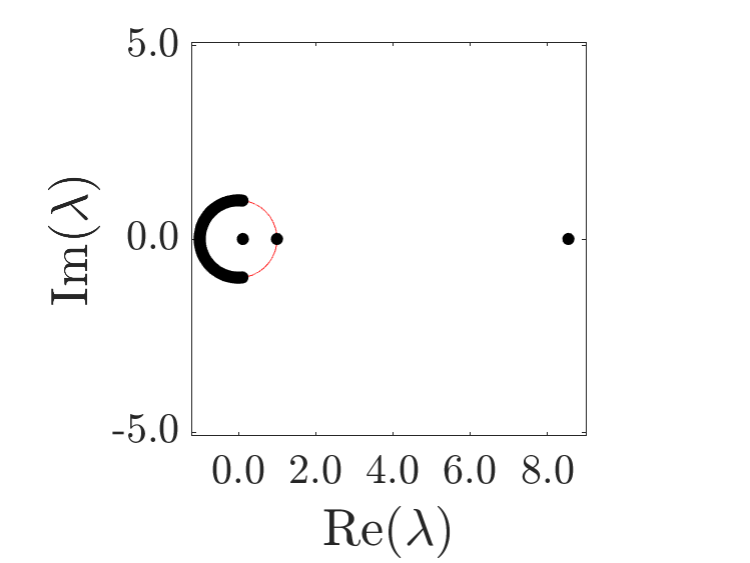} &
\includegraphics[width=.33\textwidth]{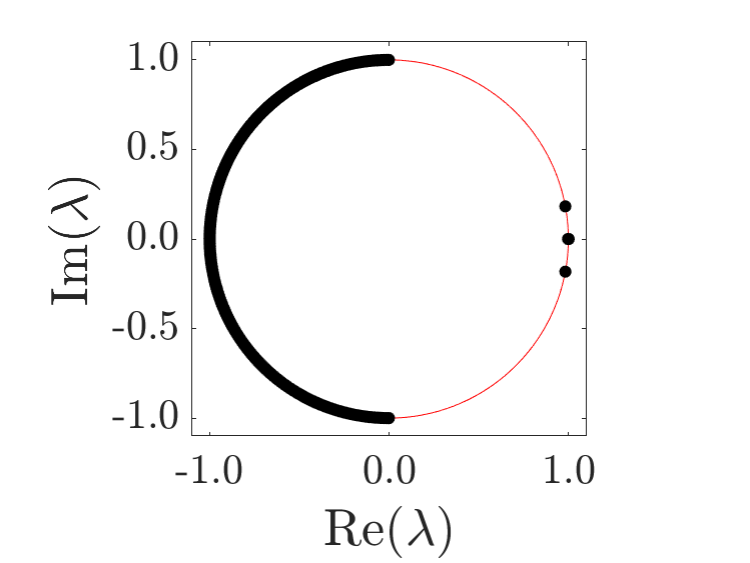}
\end{tabular}
\end{center}
\caption{Breathers with $\alpha=0.5$ and $\omega=0.8$. The top panels display the profiles of the one-site breather (left) and the inter-site breather (middle) with $C=0.5$ whereas the right panel corresponds to the out-of-phase two-site breather (also referred to as a twisted mode) with $C=0.45$.
The middle panels show the same breathers in log-log plots, so that the algebraic decay (a straight line in these plots) is better appreciated. The linear regressions have been performed by taking $n$ from $n=30$ to the end of the lattice.
The bottom panels show the Floquet spectra of the solutions over them, showing that, for the present parameter sets, the 1-site and out-of-phase 2-site breathers are linearly stable, whereas the in-phase 2-site breather is unstable.
}
\label{fig:profiles}
\end{figure}

\begin{figure}
\begin{center}
\begin{tabular}{cc}
\includegraphics[width=.5\textwidth]{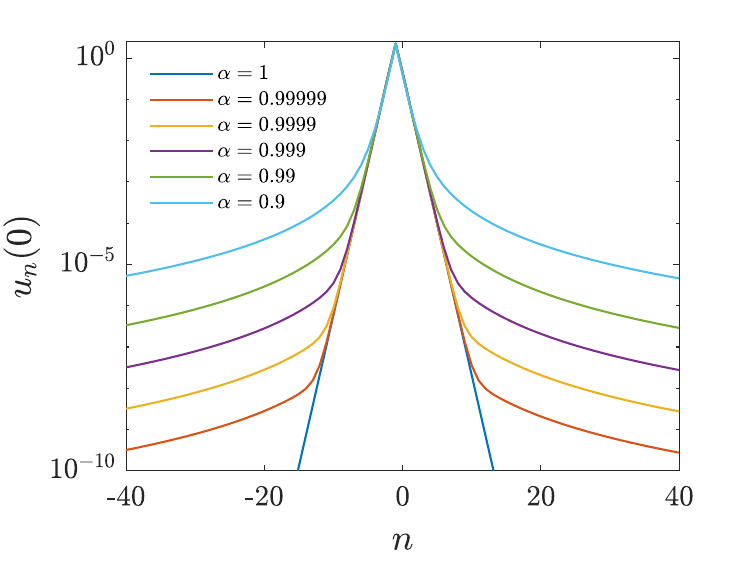} &
\includegraphics[width=.5\textwidth]{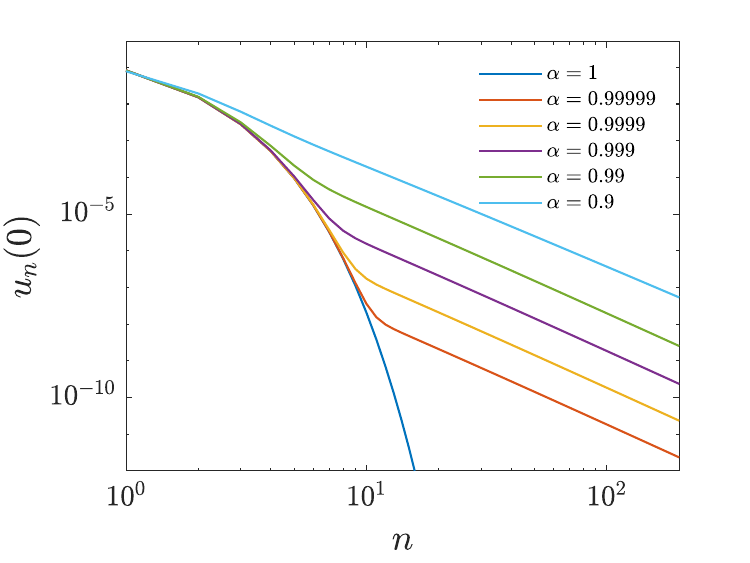} \\
\end{tabular}
\end{center}
\caption{{Profile of some selected breathers with different fractionalities $\alpha$ in semilogarithmic (left) and logarithmic (right) scales. In all the cases, $C=0.1$ and $\omega=0.8$.}}
\label{fig:decays}
\end{figure}

Before starting such a detailed analysis, we are presenting in Fig.~\ref{fig:profiles} some prototypical examples of profiles and the corresponding Floquet spectrum of discrete breathers for $\alpha=0.5$ and breather frequency $\omega=0.8$. Apart from the 1-site breather,
i.e., the quintessential onsite ground state of the system, we have shown the in-phase (the so-called inter-site mode~\cite{Martina}) and out-of-phase 2-site breathers. {Fig.~\ref{fig:decays} shows the decay in semi-logarithmic and logarithmic scales for several fractionalities ranging from $\alpha=0.9$ to $\alpha=1$. One can see that the breather decays exponentially close to the breather center, and decays algebraically from a value of $n>n_c$. The value of $n_c$ decreases with $\alpha$ and when $\alpha\leq0.9$ the decay can be practically considered to be purely algebraic. This double decay was observed for Klein-Gordon lattices with long-range interactions in the work of~\cite{Fflach} for algebraic kernels of the form $K(n)=|n|^{-s}$ when $s>3$, whereas a pure algebraic decay was observed for $1<s<3$. On the other hand, the fractional kernel (\ref{eq:kernel}) tends asymptotically to $|n|^{-s}$, with $s=1+2\alpha$ for large values of $n$ when $\alpha<1$, implying that $1<s<3$; in other words, the asymptotic decay is not fully in line with the predictions of \cite{Fflach} as there is double decay for $\alpha\gtrsim0.9$. We expect that this is caused by the dissimilarity of the two kernels for small $|n|$.}

{The decay of the tails of the on-site and inter-site breathers for $\alpha=0.5$, as shown in the middle panels of Fig.~\ref{fig:profiles},
can be well-approximated by $|n|^{-2}$, as expected from the above reasoning.}
Indeed, this decay is identified for both the onsite ground state of the
system and the inter-site (unstable, as we will see below) solution.
However, we also highlight a remarkable observation that concerns the
out-of-phase 2-site breathers. Indeed, the latter are found to decay
with a rate of $\sim|n|^{-3}$, systematically different than the one
we find for the single-peak states. {It is
straightforward to show that the asymptotic decay is, in general, $~|n|^{-(s+1)}=|n|^{-2(\alpha+1)}$, by supposing that for large $|n|$ the inter-site breather tail is the superposition of the tails of two non-interacting onsite breathers oscillating in anti-phase and separated by a lattice site, i.e., the breather decays as $~|n|^{-s}-|n+1|^{-s}$; straightforward algebra leads to the observed asymptotic dependence for the decay}.

Notice that, due to the long-range interaction introduced by the fractionality, one can expect the existence of multibreathers where the time-reversibility is broken, as  was done in \cite{Vassilis}. This case, which is precluded by the (effective reversibility) assumption made in (\ref{eq:Fourier}), is out of the scope of the present chapter, but a potentially interesting theme for further studies.

{From the Floquet spectra displayed in Fig.~\ref{fig:profiles}, one can observe the spectral stability (at least, for small coupling) of the 1-site breather and the out-of-phase 2-site breather and the instability of the in-phase 2-site breather. The linear stability/instability of these multibreather is in agreement with the predictions for low coupling of e.g. \cite{kou}.
The latter case is for nearest-neighbor coupling, hence the observation
herein is that the longer-range nature of the coupling
of the states in Fig.~\ref{fig:profiles} does not
modify their stability. Nevertheless, we will explore this issue in
further detail below (including as a function of $\alpha$, the coupling
$C$ and the breather frequency $\omega$).
On the other hand, according to \cite{DEP}, the out-of-phase 2-site breather should be nonlinearly unstable.}

\begin{figure}
\begin{center}
\begin{tabular}{cc}
\includegraphics[width=.5\textwidth]{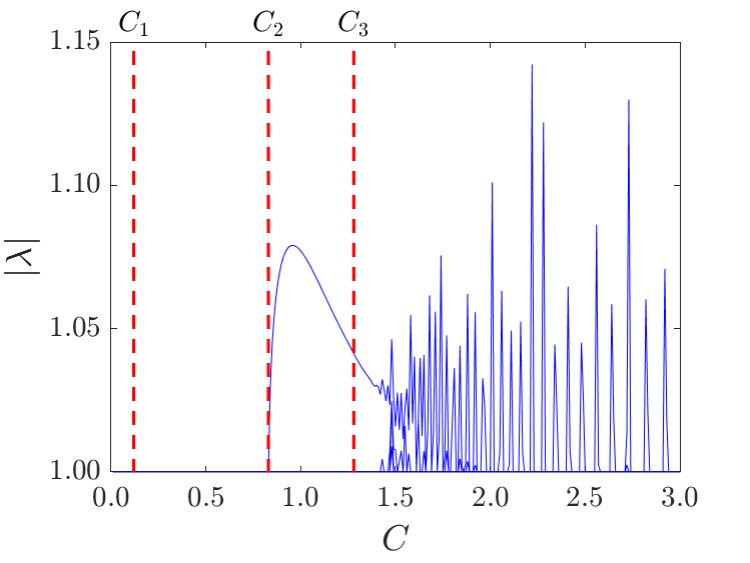} &
\includegraphics[width=.5\textwidth]{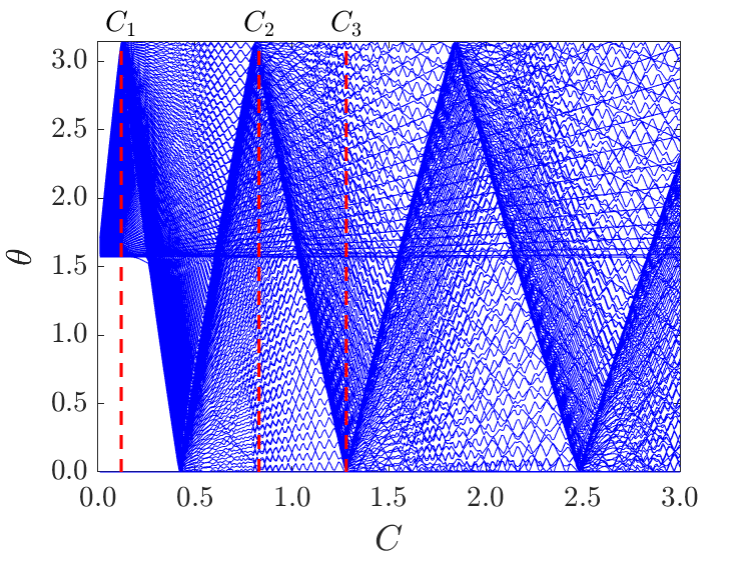} \\
\includegraphics[width=.5\textwidth]{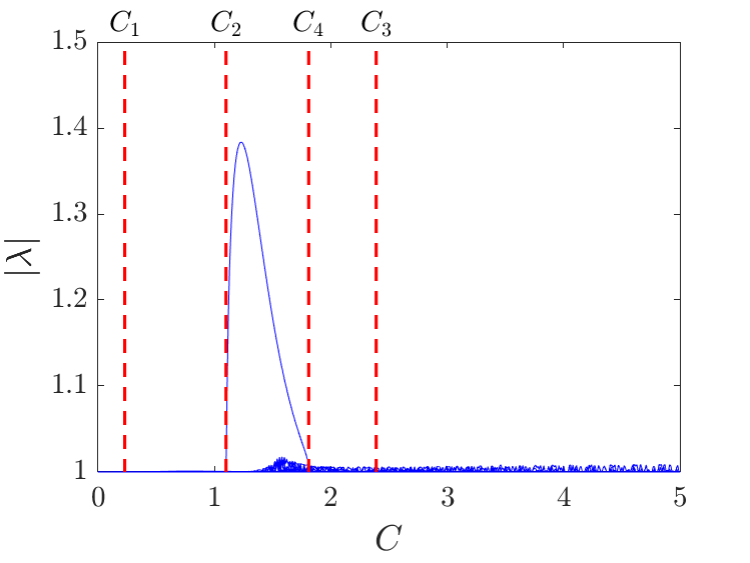} &
\includegraphics[width=.5\textwidth]{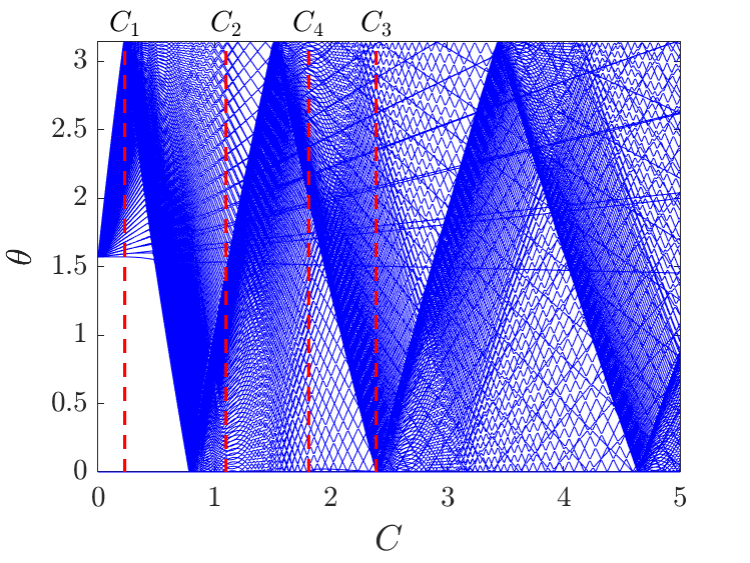} \\
\end{tabular}
\end{center}
\caption{{Dependence of the Floquet multipliers $\lambda$ with $C$ for $\omega=0.8$ and $\alpha=0.95$ (top row) and $\alpha=0.5$ (bottom row). The left panels depict the Floquet multiplier modulus, and the right panels, their argument $\theta$.} Notice that only multipliers with $\lambda\geq1$ and $\theta\geq0$ are considered, as the rest of them are trivially found as they form quadruplets. {Vertical red dashed lines indicate the critical values of $C$}.}
\label{fig:stab1}
\end{figure}

{A more detailed study of the dependence of the Floquet multipliers on the coupling constant for single-site breathers and constant frequency $\omega=0.8$ has been presented in Fig.~\ref{fig:stab1}. This figure considers {two} different values of the fractionality parameter, namely, a value close to the classical discrete Laplacian limit ($\alpha=0.95$) {and a case distant from the classical Laplacian limit ($\alpha=0.5$)}. As expected, this breather is stable when the coupling is small, and destabilizes as $C$ increases. Those destabilizations occur through three types of bifurcations. The first one that emerges when increasing the coupling, starts to occur at $C=C_1$, with {$C_1=0.12$ for $\alpha=0.95$ and $C_1=0.23$ for $\alpha=0.5$} and this particular value of frequency, are of Hopf type and caused by the overlap of phonon arcs at $\theta=\pi$ in the Floquet circle and should disappear in an infinite lattice \cite{finite}.}\footnote{{Notice that these bifurcations are so weak for $\alpha\rightarrow1$ that the modulus of multipliers responsible for the bifurcations are extremely close to $1$ and cannot be appreciated in the picture.}}
Notice that by Hopf type here, we mean instabilities associated
with quartets of complex multipliers. The dynamical manifestation
of the associated instabilities will involve both growth (due to the
departure of the relevant Floquet multiplier modulus from the
unit circle) and oscillation (due to the imaginary part of
the relevant multiplier).
{This value of $C_1$ can be analytically approximated from the following reasoning: in the case of small couplings, when the phonons are weakly hybridized with the breather, the phonon arcs in the Floquet circle are related to the phonon band $\Omega_\mathrm{ph}$ in (\ref{eq:phononband}) by $\theta_\mathrm{ph}=\pm2\pi\Omega_\mathrm{ph}/\omega\,\mathrm{mod}\,2\pi$. That is, in the anti-continuous limit, the arc collapses to a  point with $\theta=\pm2\pi/\omega$ and, in the case of $2/3<\omega<1$, the arcs expand from this point towards $\pi$ when $C$ is increased \cite{Floria}. Consequently, when $C$ is above a critical value, namely $C_1$, the two arcs collide. This occurs when the left boundary of $\theta_\mathrm{ph}$ is equal to $\pi$. In addition, because of the softness of the sine-Gordon potential, the breather frequency must be below the phonon band, i.e., $\omega<\Omega_\mathrm{ph}(0)$. As a consequence, the collision of the arcs takes place when $\omega=2\Omega_\mathrm{ph}(\pi)/3$. From (\ref{eq:phononband}), one can see that this condition transforms into:}

\begin{equation}
    \label{eq:C1}
    C_1=\frac{(9\omega^2-4)\Gamma(1+\alpha)\Gamma(\alpha)}{8\Gamma(2\alpha)\left[2F(1,-\alpha,1+\alpha;-1)-1\right]}
\end{equation}

There is a second type of bifurcation occurring at $C=C_2$, {with $C_2=0.83$ for $\alpha=0.95$ and $C_2=1.10$ for $\alpha=0.5$}, which is of exponential nature, caused by a localized {anti-symmetric} internal mode that becomes real. This bifurcation, which is the origin of moving breathers (see also the discussion in what follows), is also accompanied by a complementary bifurcation in the inter-site breather in the vicinity of $C_2$ and the emergence of a asymmetric 2-site breather \cite{Cretegny,Aubry2}. {Because of this, such a bifurcation is sometimes dubbed as {\em stability exchange
bifurcation}}. Similar bifurcations have been observed,
for instance, between the onsite and inter-site solitary
wave solutions of the saturable
discrete nonlinear Schr{\"o}dinger lattice, as discussed
in~\cite{melvin1,melvin2}.

A third kind of bifurcations, also of exponential type, firstly emerges at $C=C_3$, {with $C_3=1.28$ for $\alpha=0.95$ and $C_3=2.39$ for $\alpha=0.5$}, and is caused by the hybridization of the breather with phonons giving rise to the so-called phantom breathers \cite{phantom}. These bifurcations have their origin in the resonance of an integer multiple (which is odd because of the even spatial parity of the sine-Gordon potential{, that nullifies the terms with even $k$ in (\ref{eq:Fourier})}) of the breather frequency $\omega$ with the linear modes frequency band; actually, the value of the bifurcation point does not coincide with the resonance but it is in its vicinity.
This may have to do also with the ``density of states'' of our
numerical computation, as reflected in the precise frequencies
of the numerical representation of the continuous spectrum band.
Indeed, the hybrid (or phantom) breather can exist thanks to the
gaps in the phonon spectrum arising because of the finite lattice size. The first of those bifurcations, i.e., the one at $C=C_3$, occurs in the vicinity of the resonance of three times the breather frequency with the $q=\pi$ phonon, that is, when

\begin{equation}
    \label{eq:C3}
    C_3=\frac{(9\omega^2-1)\Gamma(1+\alpha)\Gamma(\alpha)}{2\Gamma(2\alpha)\left[2F(1,-\alpha,1+\alpha;-1)-1\right]}.
\end{equation}
{When $C$ is increased from $C_3$, there exist
(further) destabilizations and restabilizations of the breathers corresponding to resonances with phonons. They manifest as instability bubbles that can be observed in Fig.~\ref{fig:stab1}. These bubbles also appear because of the resonance between phonons emerging for $C>C_1$. The growth rate of these instabilities is larger for $\alpha=0.95$ than for $\alpha=0.5$, as it can be appreciated from the figure. In fact, the decrease of the associated growth rate when $\alpha$ decreases seems to be a general trend.}

Finally, there is a fourth bifurcation at $C=C_4>C_2$, of exponential nature, where the {eigenvalue associated to the instability arising from the bifurcation at $C_2$ enters the unit circle again}. This bifurcation is hidden amidst the bifurcations involving linear modes, especially for $\alpha$ close to 1. In that light, it is not as straightforward to identify it and we will 
consider it hereafter only for values
of $\alpha$ further away from the
discrete Laplacian limit.

{Figure~\ref{fig:stabcritical} shows the dependence of the critical values $C_1$, $C_2$ and $C_3$ on the fractionality parameter
$\alpha$. It can be observed that, when $\alpha$ is decreased from 1, all these values increase monotonically. In fact, this dependence of $C_3$ is caused by the narrowing of the phonon band
and is indeed the one most notably affected
by the fractionality, while on the other
hand the dependence of $C_1$ is much weaker
as the relevant phonon band edge collision is only
very mildly affected by the variation of $\alpha$.}
{In addition, as one can see in the middle and bottom panels of Fig.~\ref{fig:stab1} and in top panel of Fig.~\ref{fig:stab2},
the maximal modulus of the eigenvalue responsible for the stability exchange
(first occurring at $C_2$) grows when $\alpha$ decreases, and, at the same time, we observe that the instabilities emerging from the phonon hybridization weaken,
enabling the determination of $C_4$, which is equal to $1.81$ for $\alpha=0.5$.

\begin{figure}
\begin{center}
\includegraphics[width=.75\textwidth]{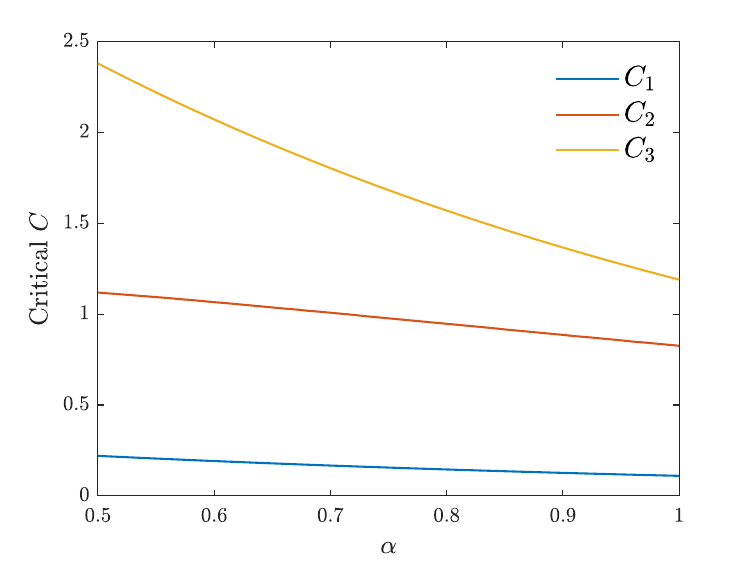}
\end{center}
\caption{Dependence of the critical values of $C$, namely $C_1$, $C_2$ and $C_3$ with respect to $\alpha$, for fixed $\omega=0.8$.}
\label{fig:stabcritical}
\end{figure}

{We have also performed a similar continuation for the 2-site in-phase multibreathers (i.e., the inter-site breather), as shown in the bottom panel of Fig.~\ref{fig:stab2}. In this case, we can see that, as expected, the breather is exponentially unstable close to the anti-continuum limit, and this exponential instability disappears at a value of $C$ close to $C_2$, observing a complementary behaviour to that of the 1-site breather in what regards to the exponential bifurcation.}
Indeed, this is strongly reminiscent
of analogous behavior in~\cite{melvin1,melvin2} and also
in the nice exposition
of~\cite{OFJ1} in the case of a
generalized discrete nonlinear
Schr{\"o}dinger model.

For the out-of-phase 2-site breather, we observe a similar behavior to the non-fractional case, with the most remarkable features being
the stability of the breathers for small $C$ and the impossibility of extending the breather to the continuum limit. Both of these features are in line with the corresponding findings in the regular Laplacian case of $\alpha=1$.
Nevertheless, it is important to highlight
that such a state features (as in the
nearest-neighbor case~\cite{DEP}) a
Floquet multiplier (point spectrum) pair whose opposite
``signature'' (see~\cite{DEP} for details)
to the continuous spectrum renders
it responsible for two features absent
for both the onsite and the in-phase,
inter-site solutions. Namely, upon
parametric variations, it is possible
for the relevant pair to collide with
the phonon band, leading to a quartet of
multipliers and a Hopf-type instability.
Secondly, even when such an instability
is absent, yet as discussed in~\cite{DEP},
a harmonic of its associated frequency
finds itself inside the phonon band,
then a {\it nonlinear} instability can
occur. We will not explore these features
in detail here, yet they may be worthwhile
of a more systematic investigation in future
work.

{At the anti-continuous limit, 2-site breathers possess two pairs of Floquet multipliers at $1+0i$. While one of the pairs will remain there because of the time-translation invariance symmetry of breathers for every value of the coupling constant, the other pair will become real when $C$ is increased from zero for the in-phase case, or become complex-conjugate (on the
unit circle) in the case of the out-of-phase breathers. The former case leads to instability and the second one to
spectral stability. It is possible to predict the dependence of the modulus (angle) of the instability (stability)-related eigenvalue with respect to $C$ when this parameter is small enough by performing an analysis similar to that of the work of~\cite{kou}. In fact, in \cite{Martina}, such an analysis was performed for in-phase 2-site breathers in the Frenkel-Kontorova model with the classical $\alpha=1$ Laplacian, but can be easily extended to out-of-phase breathers. By adapting this analysis, it is straightforward to extend those results to lattices with long-range interactions, by simply replacing $\epsilon$ by $CK^{\alpha}(1)$ in Eq.~(38) of \cite{Martina}; see the latter reference for further details. Consequently, if we express the relevant Floquet multipliers as $\lambda=\exp(i\theta)$, the angle for the above mentioned eigenvalue is given by:}

\begin{equation}\label{eq:approx2site}
    \theta=\pm\frac{2\pi}{\omega}i\sqrt{2CK^\alpha(1)\frac{\omega^2\sqrt{3(4\omega^2-1)}-(2\omega^2-1)(4\omega^2-1)}{16\omega^4-7\omega^2+1}}
\end{equation}

{If $\lambda$ is real, $\theta$ must be imaginary, fulfilling $\mathrm{Im}(\theta)=\ln(\lambda)$, and the breather is unstable. However, if $\lambda$ is complex, $\theta$ is real and the 2-site breather is
spectrally stable. Fig.~\ref{fig:stab2site} compares the dependence of $\mathrm{Im}(\theta)$ (in the former case) or $\theta$ (in the latter case) with respect to $C$ from the Floquet analysis with the approximation (\ref{eq:approx2site}). Notice that the approximation behaves worse for the out-of-phase breathers; this is caused by the limited range of existence of these multibreather structures and the existence of a bifurcation for small $C$ caused by the collision of the considered multipliers with another pair that delocalizes from the linear modes arc.}

\begin{figure}
\begin{center}
\begin{tabular}{c}
\includegraphics[width=.75\textwidth]{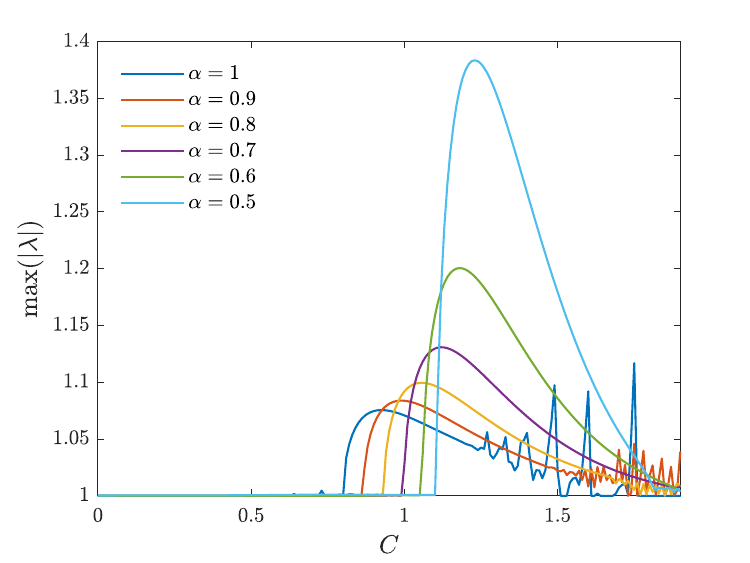} \\
\includegraphics[width=.75\textwidth]{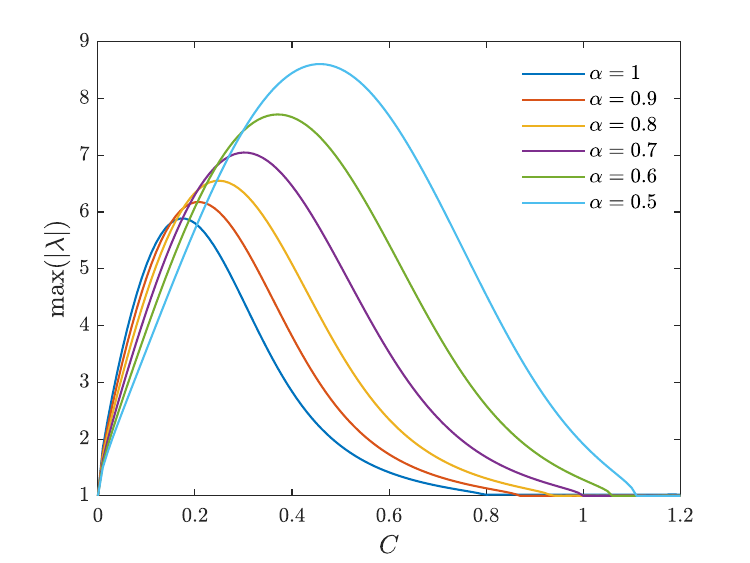}
\end{tabular}
\end{center}
\caption{Dependence of the modulus of the Floquet multipliers with the coupling $C$ for 1-site (top panel) and in-phase 2-site (bottom panel) breathers for different fractionalities when $\omega=0.8$.}
\label{fig:stab2}
\end{figure}

\begin{figure}
\begin{center}
\begin{tabular}{cc}
\includegraphics[width=0.5\textwidth]{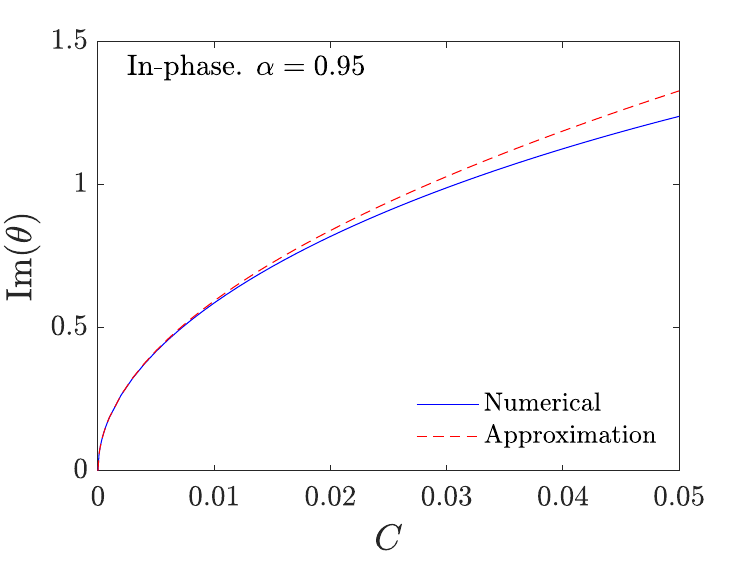} &
\includegraphics[width=0.5\textwidth]{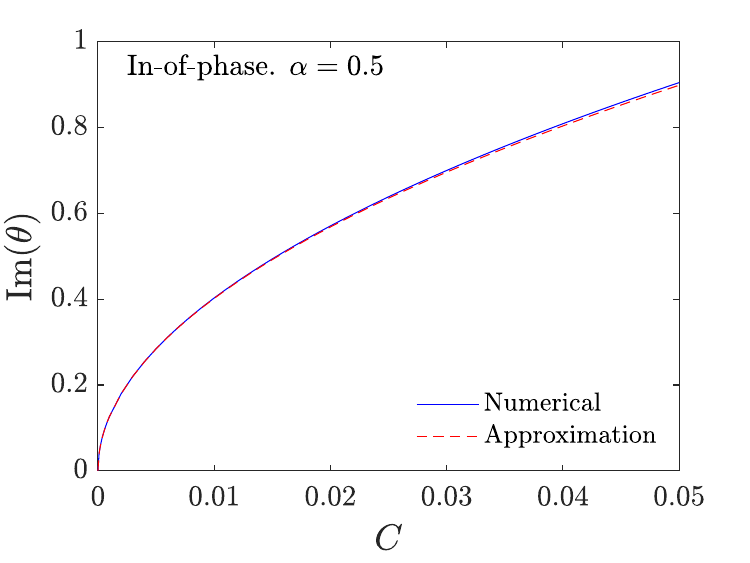} \\
\includegraphics[width=0.5\textwidth]{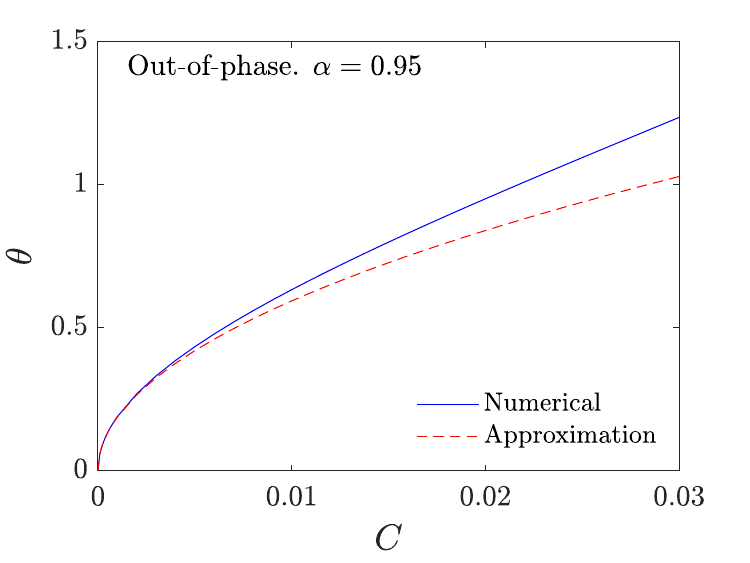} &
\includegraphics[width=0.5\textwidth]{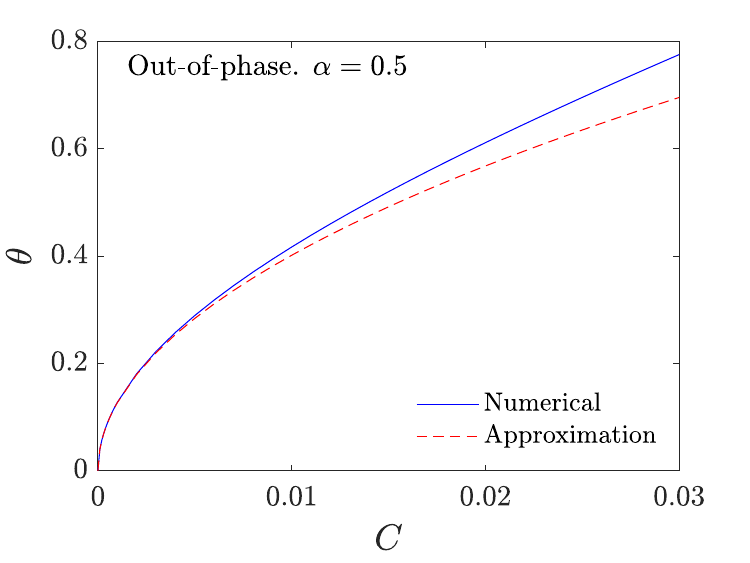}
\end{tabular}
\end{center}
\caption{{Dependence with respect to $C$ of $\mathrm{Im}(\theta)=\ln(\lambda)$ for in-phase two-site breathers (top panels) and $\theta$ for out-of-phase two-site breathers (bottom panels). In the former case, $\lambda$ corresponds to the Floquet multiplier associated with the instability of inter-site breathers (i.e. the one depicted at the bottom panel of Fig.~\ref{fig:stab2}) whereas in the latter case $\theta$ is the angle of the multiplier detaching from $\theta=0$ when the coupling is switched on. Full blue line corresponds to the numerical value and dashed red line is the approximation given by (\ref{eq:approx2site}). Left panels depict the $\alpha=0.95$ case, whereas right panels represent $\alpha=0.5$. In both cases, $\omega=0.8$.}}
\label{fig:stab2site}
\end{figure}

So far, we have chiefly concerned ourselves with the variation of the stability features of the discrete long-range breathers for different fractionalities and coupling strengths. It is also interesting to consider the dependence of the stability properties when the frequency of the obtained breather solution is varied to correlate the respective energy-frequency monotonicity with the theoretical prediction of~\cite{ourprl}~\footnote{Notice that the monotonicity of e.g., energy vs. coupling $C$ has not been associated with stability, while that of energy vs. frequency has, per the above mentioned work of~\cite{ourprl}.}. In particular, we have taken a breather at modest values of the coupling ($C=0.5$) which is stable for $\omega=0.8$ and $\alpha=0.5$ and increased its frequency. We have observed the destabilization and restabilization of the breather for frequencies in a narrow range {between $\omega\approx0.909$ and $\omega\approx0.962$}. There is a correlation of these bifurcations with a change in the monotonicity of the $H(\omega)$ dependence, as demonstrated in \cite{ourprl}. This can be observed in Figs.~\ref{fig:stabfreq} and \ref{fig:stabfreq}. A similar trend has been observed in discrete nonlinear Schr{\"o}dinger lattices with long-range interaction \cite{Gaididei}. This instability interval is absent in the classical Laplacian ($\alpha=1$) case.
Importantly, it should be highlighted
that this ``multistability'' feature
has been a principal observation of
the phenomenology of (substantial enough)
deviations from the standard discrete
Laplacian case, in the presence of
such modified dispersion, but also
in the case of modified (e.g.,
power-law) nonlinearities; see,
e.g.~\cite{malowei}.

The above obtained instability
naturally generates the question
of what the associated dynamical
evolution may be.
With that in mind, we have considered the dynamics m{in two different cases, namely a breather whose energy is close to the minimum of $H(\omega)$ and another one close to the maximum. Figure~\ref{fig:dynfreq} shows the energy density evolution for the breathers with $\omega=0.93$ and $\omega=0.961$. In the former case, the breather is detached into two moving breathers with small amplitude, whereas in the latter, the breather turns into a stable less-energetic one with frequency $0.856$, whose amplitude is modulated by a frequency $\approx0.72$. These dynamical outcomes
raise interesting additional questions for future
studies, such as the possibility to examine
systematically the potential of such lattices
to feature traveling wave states. Another
such question concerns the ``longevity'' of such
modulated (breather-on-breather) oscillations
and whether they may asymptotically decay.
{In Fig.~\ref{fig:dynfreq2} we can observe the long lifetime of the new double-frequency structure by depicting the evolution of $u_0(t)$. In addition, one can observe the profile of the new structure resembling a so-called nanopteron.}

\begin{figure}
\begin{center}
\begin{tabular}{cc}
\includegraphics[width=.5\textwidth]{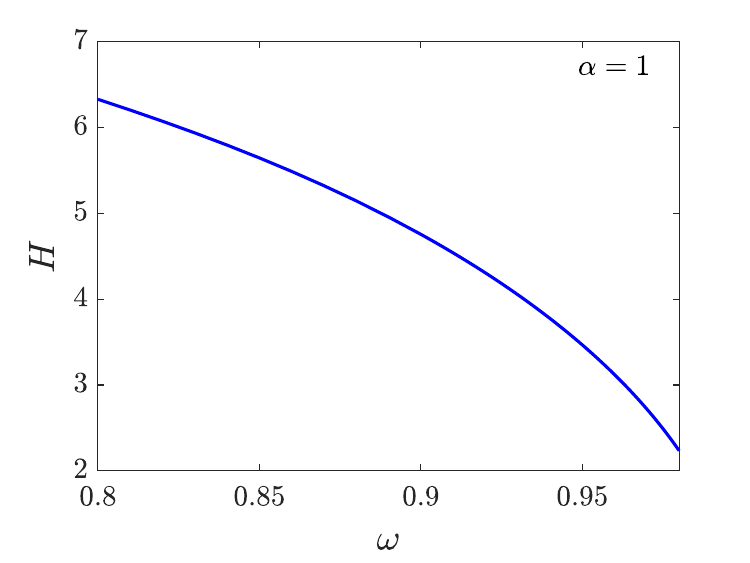} &
\includegraphics[width=.5\textwidth]{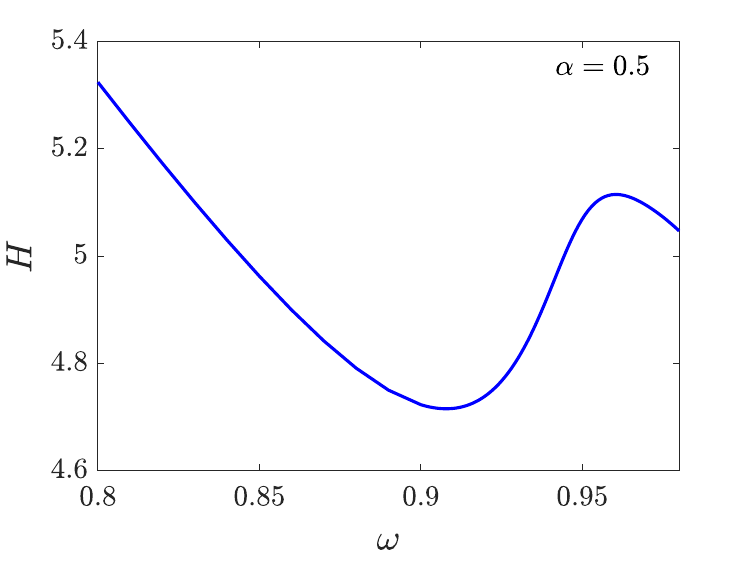} \\
\end{tabular}
\end{center}
\caption{{(Left panel) Dependence of the energy with respect to the frequency for $\omega=0.8$, $C=0.5$ and $\alpha=1$ and (right panel) $\alpha=0.5$.}}
\label{fig:stabfreq}
\end{figure}

\begin{figure}
\begin{center}
\includegraphics[width=.75\textwidth]{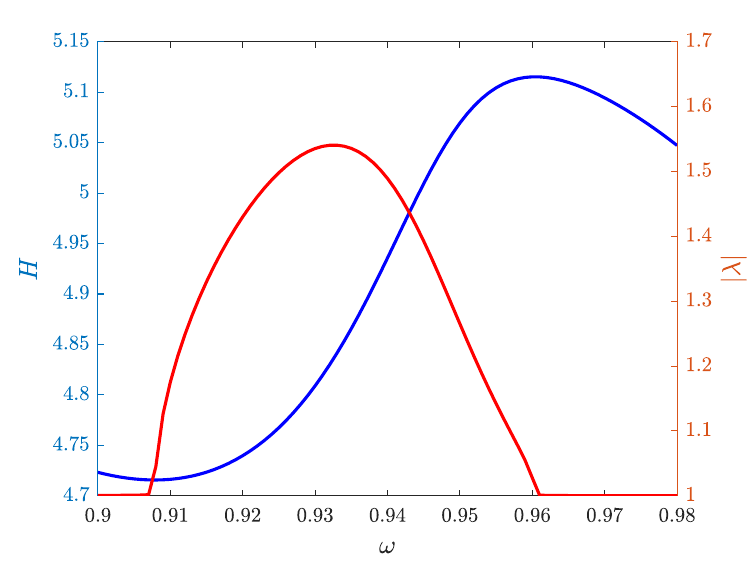} \\
\end{center}
\caption{{(Blue line) Zoom of the bottom panel of Fig.~\ref{fig:stabfreq} in the region of interest (between two extrema). (Right line) maximum modulus of the Floquet multipliers for the same frequency range of top panel. Notice how, in line with the theorem of~\cite{ourprl}, the region of instability (i.e., $|\lambda|>1$) is vertically aligned with the region of modified monotonicity (i.e., increasing dependence) of the Hamiltonian $H$ on the frequency $\omega$.}}
\label{fig:stabfreq2}
\end{figure}

\begin{figure}
\begin{center}
\begin{tabular}{cc}
\includegraphics[width=.5\textwidth]{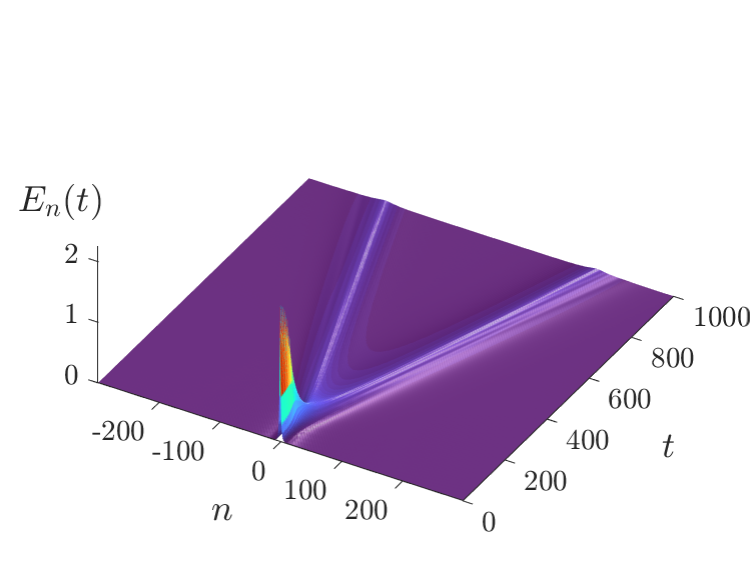} &
\includegraphics[width=.5\textwidth]{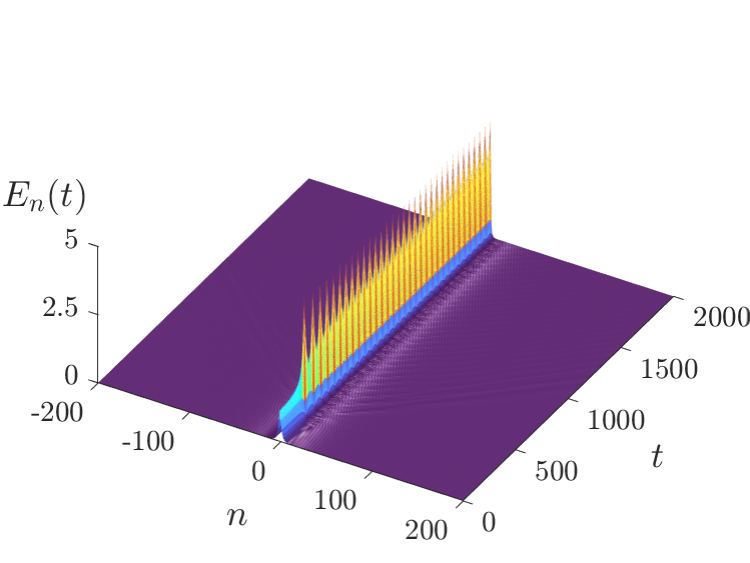} \\
\end{tabular}
\end{center}
\caption{{Evolution of the energy density of breathers with $\alpha=0.5$, $C=0.5$, $\omega=0.8$ and $\omega=0.93$ (left panel) or $\omega=0.961$ (right panel). The initial condition of the simulations has been attained by perturbing the stationary breather in the direction of he unstable eigenmode.}}
\label{fig:dynfreq}
\end{figure}

\begin{figure}
\begin{center}
\begin{tabular}{cc}
\includegraphics[width=.5\textwidth]{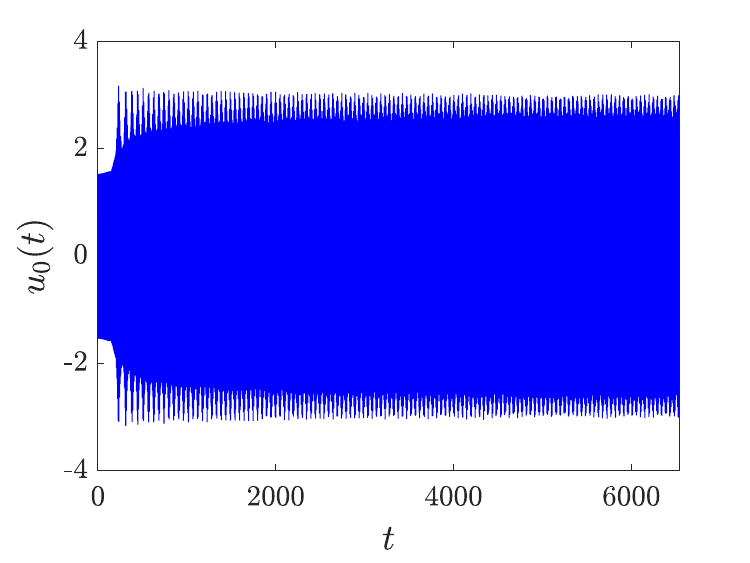} &
\includegraphics[width=.5\textwidth]{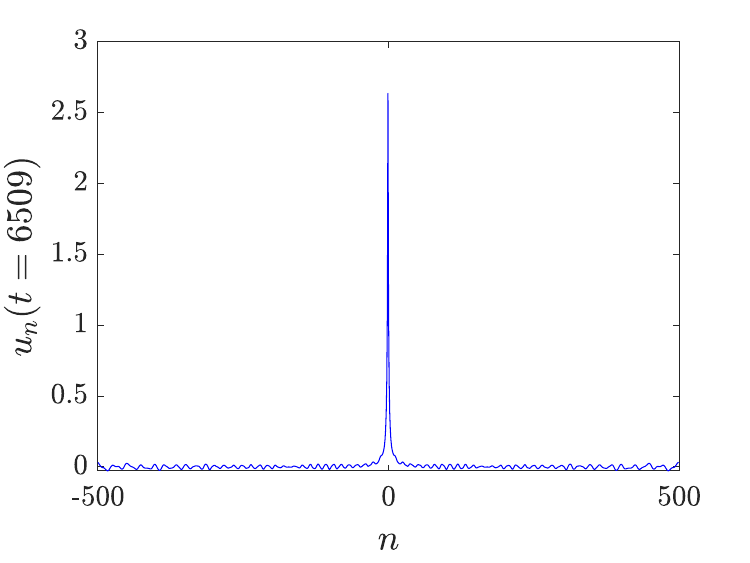} \\
\end{tabular}
\end{center}
\caption{{(Left panel) Evolution of the oscillations in the central site in the dynamics of the perturbed breather whose energy density was displayed in the right panel of Fig.~\ref{fig:dynfreq}. The profile of the breather at one of the final times of the simulations is displayed in the right panel.}}
\label{fig:dynfreq2}
\end{figure}

To end this section, we briefly touch upon moving breathers. These are attained by perturbing a discrete 1-site breather in the $[C_2,C_4]$  interval along the eigenvector associated with the real Floquet multiplier. {Recall that in this interval 1-site breathers are exponentially unstable because of a localized eigenmode that is associated to a multiplier larger than one; in addition, such a (translational) eigenmode is spatially anti-symmetric.} Upon perturbing such unstable breathers by kicking the corresponding unstable eigendirection,
Fig.~\ref{fig:moving} displays the evolution of the energy density $E_n$ of the resulting moving breathers with $\alpha=0.95$ and $\alpha=0.5$ for {a value of $C$ close to (and higher than) $C_2$. We can observe that the motion of the breather becomes decelerated in the latter case, as there seems to be a high amount of emitted radiation. A possible explanation of this last phenomenon can rely on the fact that the tails have a higher amplitude when $\alpha$ is decreasing and, consequently, when the tails are perturbed there will be more radiation emission than if their amplitude were smaller. A similar trend was observed in a Klein-Gordon chain with Morse substrate potential and a long range interaction with kernel $K(n)=-|m|^{-3}$ \cite{JesusDNA}. }

\begin{figure}
\begin{center}
\begin{tabular}{cc}
\includegraphics[width=.5\textwidth]{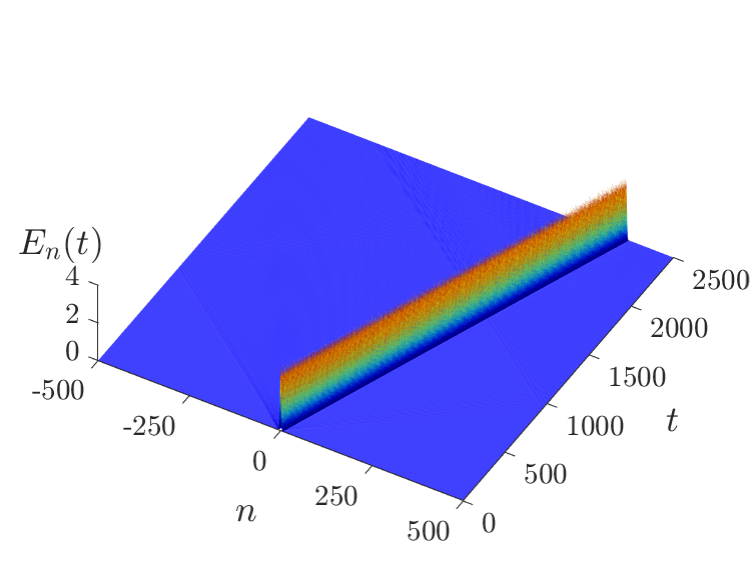} &
\includegraphics[width=.5\textwidth]{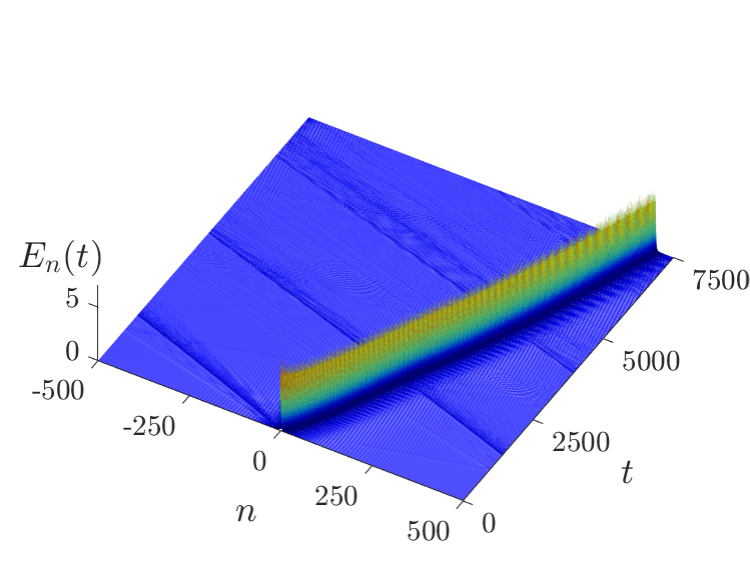}
\end{tabular}
\end{center}
\caption{Evolution of the energy density for moving breathers with $\omega=0.8$ and (left panel) $C=0.81$ and $\alpha=0.95$, and (right panel) $C=1.10$ and $\alpha=0.5$.}
\label{fig:moving}
\end{figure}

\section{Conclusions and Future Challenges}

In the present work we have explored the intersection between a prototypical lattice nonlinear dynamical model, i.e. the discrete sine-Gordon equation (or Frenkel-Kontorova model) and the emerging notion of fractional systems, which at the discrete level amounts to a long-range interaction that decays with a suitable power-law, as discussed above. Within this setting, we have explored a central coherent
structure, generic within such lattices, namely the possibility of formation of discrete breathers. We have found that
such states continue to exist in the typical waveforms anticipated from prior experience with nearest-neighbor (or similar) models. However, the structural characteristics of the associated tails are dictated by the presence of the long-range interactions and indeed feature power law decay
of the breather tails. We have also explored the spectral stability of such breather states by computing the associated monodromy matrix and accordingly evaluating the respective Floquet multipliers. We thus identified both the real (exponential in time) and the complex (oscillatorily growing in time) multiplier settings and quantified the corresponding instability rates, when appropriate. We used monoparametric
continuations both in the strength of the coupling, showing how the corresponding continuum (fractional derivative) limit is approached, as well as in frequency, connecting the existence results and their energy-frequency diagrams with the stability ones through the monotonicity of such diagrams.
We examined the relevant scenarios
for different values of the fractionality
parameter $\alpha$,
both
in the vicinity of the nearest neighbor
$\alpha=1$ limit, and further away
from it.
Finally, we also examined the dynamical outcome of a suitable kick on the identified discrete breathers and found that, under suitable conditions, the breathers can be boosted to a moving state that persists over the long-time dynamical simulations that we performed. This was in line with the
stability exchange results that we
found (as, e.g. the coupling was varied)
between onsite and inter-site solutions,
as has been previously observed in
different models. Also, the long-range
nature of the interaction was
found to lead to multi-stability for
suitable values of the fractional and
coupling
parameters, in combination with
ranges of the breather frequency.

Naturally, this study paves the way for further explorations in the future. One aspect that is of particular interest
involves the feature that discrete breathers do not persist at the continuum limit. We expect that this feature
of the standard Laplacian (continuum) model is still valid for fractional continuum models, but a further examination
of this theme would be relevant to consider. Should such fractional continuum breathers be absent, an analysis
of how the discrete long-range breathers terminate as the fractional derivative continuum limit is approached,
would be relevant to perform.
In our studies, the chosen values for the
onsite breather and the in-phase 2-site (inter-site) breather, namely $C=5$, suggests that these structures exist fairly close to the corresponding continuum
limit and hence the corresponding exploration of
these states in the continuum limit is of particular
interest; see~\cite{Alfimov} for a relevant discussion.
On the contrary, as
was expected (from the nearst-neighbor limit), the out-of-phase 2-site breather cannot be continued up to the continuum limit,
given the inherently discrete nature of such a state.

An additional item worthwhile to study
even in the one-dimensional realm,
as indicated above, concerns the potential
of genuinely traveling discrete breathers
in fractional discrete nonlinear media, {in
a way similar to what
was done, e.g., in \cite{duran} for a
nearest-neighbor setting}.
Of particular interest also, as demonstrated
above, are the out-of-phase multi-site
discrete breathers that possibly also
merit a separate investigation.
{Additionally, recall that the settings
considered herein were somewhat restricted
in terms of the fractionality parameter to
values more proximal to the nearest neighbor
limit (such as $0.5< \alpha< 1$). It would
be of interest to seek to address the
highly nonlocal interaction regime for
$\alpha<0.5$.}
Finally, in a spirit similar to the earlier work of~\cite{m10} and especially of the
more recent considerations of~\cite{m21}, it would be of relevance to explore such discrete long-range interaction models beyond one spatial dimension, considering therein the more elaborate possibility of states bearing vorticity, i.e., a phase-winding by multiples
of $2 \pi$ over a discrete contour. Such studies are currently under consideration and will be accordingly
reported in future publications.

\begin{acknowledgement}
This material is based upon work supported by the US National Science Foundation under Grants PHY-2110030 and DMS-2204702 (P.G.K.). J.C.-M. acknowledges support from EU (FEDER program 2014-2020) through MCIN/AEI/10.13039/501100011033 under the project PID2020-112620GB-I00.
\end{acknowledgement}

\end{document}